\documentclass[conference]{IEEEtran}

\IEEEoverridecommandlockouts

\usepackage{cite}
\usepackage{amsmath,amssymb,amsfonts}
\usepackage{algorithmic}
\usepackage{graphicx}
\usepackage{textcomp}
\usepackage{xcolor}
\usepackage{array}
\usepackage{pifont}
\usepackage[english]{babel}

\def\BibTeX{{\rm B\kern-.05em{\sc i\kern-.025em b}\kern-.08em
    T\kern-.1667em\lower.7ex\hbox{E}\kern-.125emX}}
\begin{document}

\begin{titlepage}

     \vspace{1cm}
        Full Citation: J. Guan and A. Morris, "Design Frameworks for Hyper-Connected Social XRI Immersive Metaverse Environments," in IEEE Network, vol. 37, no. 4, pp. 12-21, July/August 2023, doi: 10.1109/MNET.005.2300129.
keywords: {Metaverse;Corporate acquisitions;Mixed reality;Switches;User experience;Digital twins;Internet of Things;Extended reality},

       \vspace*{1cm}

       \copyright2024 IEEE. Personal use of this material is permitted.  Permission from IEEE must be obtained for all other uses, in any current or future media, including reprinting/republishing this material for advertising or promotional purposes, creating new collective works, for resale or redistribution to servers or lists, or reuse of any copyrighted component of this work in other works.

       \vspace{1.5cm}
  
\end{titlepage}


\title{Design Frameworks for Hyper-Connected Social XRI Immersive Metaverse Environments
\thanks{Tri-council of Canada, Canada Research Chairs Program.}
}

\author{\IEEEauthorblockN{Jie Guan, Alexis Morris}
\IEEEauthorblockA{\textit{Adaptive Context Environments Lab} \\
\textit{OCAD University}\\
Toronto, Canada \\
\{jie.guan, amorris\}@ocadu.ca}








}


\maketitle

\begin{abstract} 
The metaverse refers to the merger of technologies for providing a digital twin of the real world and the underlying connectivity and interactions for the many kinds of agents within. As this set of technology paradigms — involving artificial intelligence, mixed reality, the internet-of-things and others — gains in scale, maturity, and utility there are rapidly emerging design challenges and new research opportunities. In particular is the metaverse disconnect problem, the gap in task switching that inevitably occurs when a user engages with multiple virtual and physical environments simultaneously. Addressing this gap remains an open issue that affects the user experience and must be overcome to increase overall utility of the metaverse. This article presents design frameworks that consider how to address the metaverse as a hyper-connected meta-environment that connects and expands multiple user environments, modalities, contexts, and the many objects and relationships within them. This article contributes to i) a framing of the metaverse as a social XR-IoT (XRI) concept, ii) design Considerations for XRI metaverse experiences, iii) a design architecture for social multi-user XRI metaverse environments, and iv) descriptive exploration of social interaction scenarios within XRI multi-user metaverses. These contribute a new design framework for metaverse researchers and creators to consider the coming wave of interconnected and immersive smart environments.

\end{abstract}


\section{Introduction} \label{Introduction}

\begin{figure*}[tbh]
 \centering 
 \includegraphics[width=\linewidth]{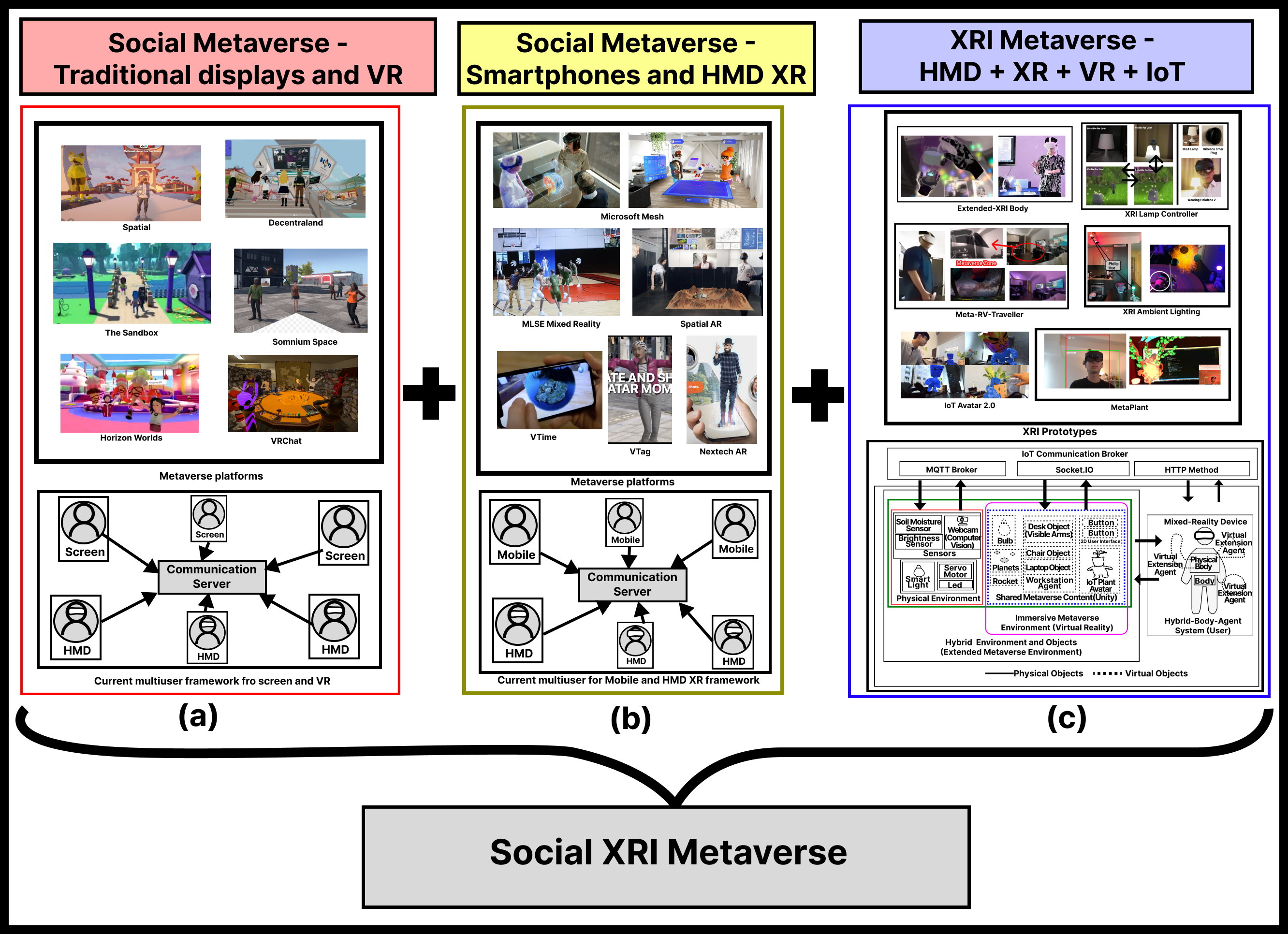}
 \caption{Existing social metaverse platforms focus on (a) screen and VR experiences and (b) mobile mixed reality and head-mounted experiences. This can be combined with (c) the XRI concept toward a new social-XRI metaverse. Table \ref{Comparison} shows more detail and comparison of the systems shown above.}
 \label{SocialXRIBackground}
\end{figure*}

The metaverse can be described as ``a hypothetical synthetic environment linked to the physical world'' \cite{lee2021all} and, this concept has been growing in terms of the underlying technologies that support metaverse eco-systems. These technologies include extended reality, artificial intelligence, Internet of Things, cloud computing, blockchains and others, each of which has become mainstream productive application domains \cite{lee2021all}. The merger of these gives rise to the visions of the metaverse presented in early conceptualizations; toward a virtual space that is persistent and pervasive, portals between the real world and the virtual world in immersive and seamless ways. The metaverse concept has also been considered as the next-generation Internet, with many high-tech companies engaging in this area to build the infrastructure\cite{cheng2022will} and gain access to new opportunities and use cases. As the paradigm of the metaverse and its underlying technologies grows in adoption, there are also new research questions and challenges that must be addressed to enable it to reach its full potential as a new interface and medium of communication. Further, in the context of multi-user metaverse environments, there are new human factors that emerge at the physical, psychological, social, organizational, and political levels where a human-tech approach \cite{vicente2013human} is needed to ensure that shared immersive metaverse environments become reliable, safe, and effective spaces for social interactions.

The current metaverse, with characteristics related to ``perpetual, shared, concurrent, and 3D virtual spaces concatenating into a perceived virtual universe'' \cite{lee2021all} brings with it a naturally occurring gap between virtual and physical environments\cite{guan2022metabuild}. To cope with this disconnect between the real and virtual worlds, richer connections are needed to create an immersive and hyper-connected spatial experience for users. As in the authors' previous work\cite{guan2022metabuild}, addressing the Metaverse disconnect problem requires approaches to hyper-connect the user, the virtual, and the physical environment by making hybrid virtual-physical objects using XR-IoT (XRI) design frameworks. To date, these have focused on single-user experiences; however, to address the complex practical nature of metaverse relationships, a focus is needed on bringing multiple users into the same hyper-connected experience. 
At present, such a framework is not common to the metaverse platforms available, although there are works toward this direction, as shown in Table \ref{Comparison}. To advance the research in this direction, this work explores how to design a framework for multi-user shared hyper-connected extended metaverse immersive smart environments. This framework builds on the XRI and extended metaverse frameworks as stated in \cite{guan2022metabuild}, which focus on single-user scenarios, and extends these toward multi-user designs.
This article contributes to i) a framing of the metaverse as a social XR-IoT (XRI) concept, ii) design Considerations for XRI metaverse experiences, iii) a design architecture for social multi-user XRI metaverse environments, and iv) descriptive exploration of social interaction scenarios within XRI multi-user metaverses. This extends the XRI concept from single-user to multi-user scenarios, provides dimensions for designing such systems and their core components, and examines the complex relationships between users and other agents within XRI multi-user metaverses. 

These contributions are presented as follows: Section \ref{Introduction}  has provided a motivation toward multi-user XRI social metaverses. Section \ref{TowardScialXRIMetaverse} presents existing social metaverse frameworks and their properties, showing the opportunity to merge these with XRI concepts as a social XRI metaverse. Section \ref{DesignConsideration} highlights the design dimensions and underlying decisions required for creating XRI metaverse experiences. Section \ref{DesignArchitecture} synthesizes these design dimensions into a new architecture for social multi-user XRI metaverse environments. Section \ref{SocialInteractionScenarios} presents an exploration of the kinds of interaction scenarios that users within an XRI multi-user metaverse environment will experience, including agents, avatars, and environment objects. Section \ref{Discussion} provides a closing discussion and Section \ref{Summary} presents a summary.

\section{Toward a Social XRI Metaverse} \label{TowardScialXRIMetaverse}


Table \ref{Comparison} presents the comparison of the various features of social metaverse with the traditional displays and VR platforms  (see Figure \ref{SocialXRIBackground}(a)), social metaverse with smartphones and HMD XR (see Figure \ref{SocialXRIBackground}(b)), and XRI prototypes  (see Figure \ref{SocialXRIBackground}(c)). 

Social metaverses designed for traditional displays and VR platforms (i.e., the most common form of metaverse design at present) typically focus on providing an immersive virtual experience for users to connect and interact with each other remotely. However, these platforms often do not have the capability of integrating physical space into the virtual experience, as they are not IoT-enabled. As a result, these existing approaches are limited to scanning the user environment and anchoring content. However, they are limited in terms of their ability also to obtain detailed information about objects and IoT edge devices that may be active in the user's environment.  

Social metaverse with smartphones and HMD XR applications provide a mixed-reality experience with users remotely and virtually present via XR devices and displays. The XRI prototypes, as in Table \ref{Comparison},  are equipped with IoT capabilities, allowing them to interact with the physical space in a more meaningful way, \cite{morris2021xri}. However, these prototypes often lack the ability to support multi-user experiences, which is a critical aspect of social VR and metaverse platforms. The lack of multi-user support in XRI prototypes can limit the level of interaction and collaboration between users in a shared virtual environment.

\subsection{Related Work} 
\label{relatedwork}
\subsubsection{Online Social Environments and Social Virtual Reality}

Social virtual reality (VR) platforms provide shared connectedness, and immersive virtual environments where users can interact and socialize with each other\cite{cheng2022we}. A social metaverse space is not only presented in VR platforms but can also involve traditional two-dimensional displays, such as Decentraland and Spatial (see Table \ref{Comparison}). Various events are hosted in the metaverse to enable the social values; for example, LNY metaverse\footnote{https://news.yahoo.com/interview-karen-x-cheng-her-220250345.html (accessed on 05-February-2023)} created a virtual space and hosted an event in spatial to celebrate the Lunar New Year.


\subsubsection{Social XR Metaverse}
The Social XR metaverse (see Figure \ref{SocialXRIBackground}(b)) refers to multi-user mixed reality local experiences, i.e., with each user in the same room seeing the same virtual objects (see Microsoft Mesh in Table \ref{Comparison}), or remote experiences, i.e., with users present remotely as a virtual avatar, such as in Spatial-AR \footnote{https://www.wired.com/story/spatial-vr-ar-collaborative-spaces/ (accessed on 07-February-2023)}. The Digital Labs of MLSE (Maple Leaf Sports and Entertainment) are working on Mixed Reality for viewing the experiences of NBA and NHL fans. It is an example of a multi-user watching the same virtual experience for sports and entertainment. Regarding handheld mobile AR multi-user experience, VTime XR mode could visualize remote users and interact with them synchronously. At the same time, Vtag could anchor a user's avatar on a location that could then be visualized by other people in AR asynchronously. Figure \ref{SocialXRIBackground} highlights how these different configurations of the metaverse have the opportunity to be combined.

\subsubsection{XRI Extended Metaverse}
The Concept of XRI has been addressed in \cite{morris2021xri}, which is the hybridization of XR and IoT that aims to enhance the connection between virtual and physical objects and the environment. As the metaverse is becoming mainstream in recent years, the concept of XRI is being applied to extend the metaverse to physical spaces through IoT with multiple prototypes, as in \cite{guan2022metabuild}\cite{guan2022thesis}. The extended-XRI body is introduced in \cite{guan2022extendedbody} with virtual body extension for users to enhance the human-in-the-loop hyper-connected metaverse environment.

\subsection{Defining Characteristics of Social XRI Metaverse Environments}

\begin{table*}[h!]
\begin{minipage}{\textwidth}
\centering
\begin{tabular}{ | m{5em} | m{0.5cm}| m{0.2cm} | m{0.2cm} |m{1.1cm} |m{0.3cm} |m{0.6cm} |m{1cm} |m{1cm} |m{1.1cm} |m{1.3cm} | m{0.67cm} |m{1.26cm}| m{1.41cm}|} 
\hline
Platforms & Social & VR & XR &  Traditional display& IoT &  IoT Avatar & Local Environment & Remote Environment & Blockchain & Avatarization & Agency & Synchronous & Asynchronous\\
  \hline
  Horizon Worlds\footnote{https://www.oculus.com/horizon-worlds/  (accessed on 05-February-2023)}{\value{Horizon}}
& \ding{51} & \ding{51} & \ding{55}& \ding{55}& \ding{55}& \ding{55}& \ding{55}& \ding{51}& \ding{55}& \ding{51} & \ding{55}& \ding{51}& \ding{55}\\ 
  \hline
  
  Spatial\footnote{https://www.spatial.io/  (accessed on 05-February-2023)}& \ding{51} & \ding{51} & \ding{55} & \ding{51}& \ding{55}& \ding{55}& \ding{55}& \ding{51}& \ding{55} & \ding{51}& \ding{55}& \ding{51}& \ding{55}\\ 
  \hline
  Decentraland\footnote{https://decentraland.org/  (accessed on 05-February-2023)} & \ding{51} & \ding{55} & \ding{55} & \ding{51}& \ding{55}& \ding{55}& \ding{55}& \ding{51}& \ding{51} & \ding{51}& \ding{55}& \ding{51}& \ding{55}\\ 
  \hline
  The Sandbox\footnote{https://www.sandbox.game/  (accessed on 05-February-2023)} & \ding{51} & \ding{55} & \ding{55} & \ding{51}& \ding{55}& \ding{55}& \ding{55}& \ding{51}& \ding{51} & \ding{51}& \ding{55}& \ding{51}& \ding{55}\\ 
  \hline
  Somnium Space\footnote{https://somniumspace.com/  (accessed on 05-February-2023)} & \ding{51} & \ding{51} & \ding{55} & \ding{51}& \ding{55}& \ding{55}& \ding{55}& \ding{51}& \ding{51} & \ding{51}& \ding{55}& \ding{51}& \ding{55}\\ 
  \hline
Microsoft Mesh\footnote{https://www.microsoft.com/mesh  (accessed on 05-February-2023)}& \ding{51} & \ding{55} & \ding{51} & \ding{51} &\ding{55} & \ding{55} & \ding{51} & \ding{51} & \ding{55}  & \ding{51}& \ding{55}& \ding{51}& \ding{55}\\ 
  \hline
  MLSE Raptors Demo\footnote{https://www.thestar.com/sports/raptors/2023/01/24/the-future-of-sports-mixed-reality-viewing-experiences-coming-for-nhl-nba-fans.html (accessed on 08-February-2023)}& \ding{51} & \ding{55} & \ding{51} & \ding{51} &\ding{55} & \ding{55} & \ding{51} & \ding{55} & \ding{55}  & \ding{51}& \ding{55}& \ding{51}& \ding{55}\\ 
  \hline
  vTime XR - AR Mode\footnote{https://vtag.com/  (accessed on 06-February-2023)} & \ding{51} & \ding{55} & \ding{51} & \ding{51}& \ding{55}& \ding{55}& \ding{55}& \ding{51}& \ding{55} & \ding{51}& \ding{55}& \ding{51}& \ding{55}\\ 
\hline
  VTag\footnote{https://vtag.com/  (accessed on 06-February-2023)} & \ding{51} & \ding{55} & \ding{51} & \ding{51}& \ding{55}& \ding{55}& \ding{55}& \ding{51}& \ding{55} & \ding{51}& \ding{55}& \ding{51}& \ding{51}\\ 
  \hline

  Nextech AR - ARway\footnote{https://www.nextechar.com/arway  (accessed on 06-February-2023)} & \ding{51} & \ding{55} & \ding{51} & \ding{51}& \ding{55}& \ding{55}& \ding{55}& \ding{51}& \ding{55} & \ding{51}& \ding{55}& \ding{55}& \ding{51}\\ 
  \hline

  IoT Avatar 2.0\cite{morris2020toward} & \ding{55} & \ding{55} & \ding{51} & \ding{55}& \ding{51}& \ding{51}& \ding{51}& \ding{55}& \ding{55} & \ding{55}& \ding{51}& \ding{51}& \ding{55}\\ 
  \hline

  XRI Workstation\cite{morris2021xri} & \ding{55} & \ding{55} & \ding{51} & \ding{55}& \ding{51}& \ding{55}& \ding{51}& \ding{55}& \ding{55} & \ding{55}& \ding{55}& \ding{51}& \ding{55}\\ 
  \hline

  XRI Metaverse Prototypes \cite{guan2022metabuild}\cite{guan2022thesis} & \ding{55} & \ding{51} & \ding{51} & \ding{55}& \ding{51}& \ding{55}& \ding{51}& \ding{55}& \ding{55} & \ding{55}& \ding{55}& \ding{51}& \ding{55}\\ 
  \hline

  XRI Body\cite{guan2022extendedbody}& \ding{55} & \ding{55} & \ding{51} & \ding{55}& \ding{55}& \ding{51}& \ding{51}& \ding{55}& \ding{55} & \ding{51}& \ding{55}& \ding{51}& \ding{55}\\ 
  \hline

  Proposed Social XRI Metaverse & \ding{51} & \ding{51} & \ding{51} & \ding{51}& \ding{51}& \ding{51}& \ding{51}& \ding{51}& \ding{51} & \ding{51}& \ding{51}&\ding{51}& \ding{51}\\ 
  \hline
\end{tabular}
\caption{Example Social Metaverse Applications, Extended Metaverse XRI Prototypes, and Criteria for Social XRI experiences. \\ Key: \ding{51} meets criteria; \ding{55} does not meet criteria.}
 \label{Comparison}
 \end{minipage}
\end{table*}

Online Social Environments, as seen in Table \ref{Comparison}, are growing more popular. These social virtual spaces and XRI spaces can be described according to multiple characteristics, and the list below highlights key criteria the metaverse system ideally would address:
\begin{itemize}
    \item \textbf{Social} - Can allow multiple users to interact with each other using different modalities (text, speech, avatars, gestures, images, videos, etc).
    \item \textbf{Virtual Reality} - Can allow users to experience virtual immersive environments.
    \item \textbf{Mixed Reality (XR)} - Can allow users to experience virtual objects overlaid in the real-world environment.
    \item \textbf{Traditional 2D Displays and Screens} - Can allow users to experience virtual content without requiring an immersive head-mounted display. 
    \item \textbf{Internet-of-Things (IoT)} - Can allow objects in the user's environment to sense and react to changes and also share information across online networks. 
    \item \textbf{IoT Avatar}\cite{morris2020toward} - Can allow objects in the user's environment to interact using virtual and mixed reality avatars.
    \item \textbf{Local Environment} - Refers to the specific environment location where a user is physically present (local user, \cite{almeida2022telepresence}). 

    \item \textbf{Remote Environment} - Refers to the specific environment location where a user is remote or telepresent (remote user \cite{almeida2022telepresence}). 

    \item \textbf{Blockchain} - Online environment platform is integrated with or supported by blockchain technology \cite{lee2021all}.

    \item \textbf{Avatarization}\cite{genay2022Avatarization} \cite{guan2022extendedbody} - Refers to the virtual representation and embodiment of a user, or an object.

    \item \textbf{Agency}\cite{holz2011mira}\cite{wooldridge1995agent} - Refers to the ability of users and digital agents to perceive the environment (local or remote; virtual or physical) and take actions within the environment (local or remote; virtual or physical). 

    \item \textbf{Synchronous} - Refers to the continuous real-time interactions between users and other agents within the environment (local or remote; virtual or physical).

    \item \textbf{Asynchronous} - Refers to the non-continuous communication interactions between users and other agents within the environment (local or remote; virtual or physical).

       
\end{itemize}



\section{Design Considerations for an XRI Metaverse Experience}\label{DesignConsideration}

In order to address the social design challenges of multi-user and multi-agent XR-IoT metaverse systems, this work first highlights the design components for XRI systems, and later extends these toward multi-user and multi-agent scenarios.


\begin{figure*}[!htb]
 \centering 
 \includegraphics[width=\linewidth]{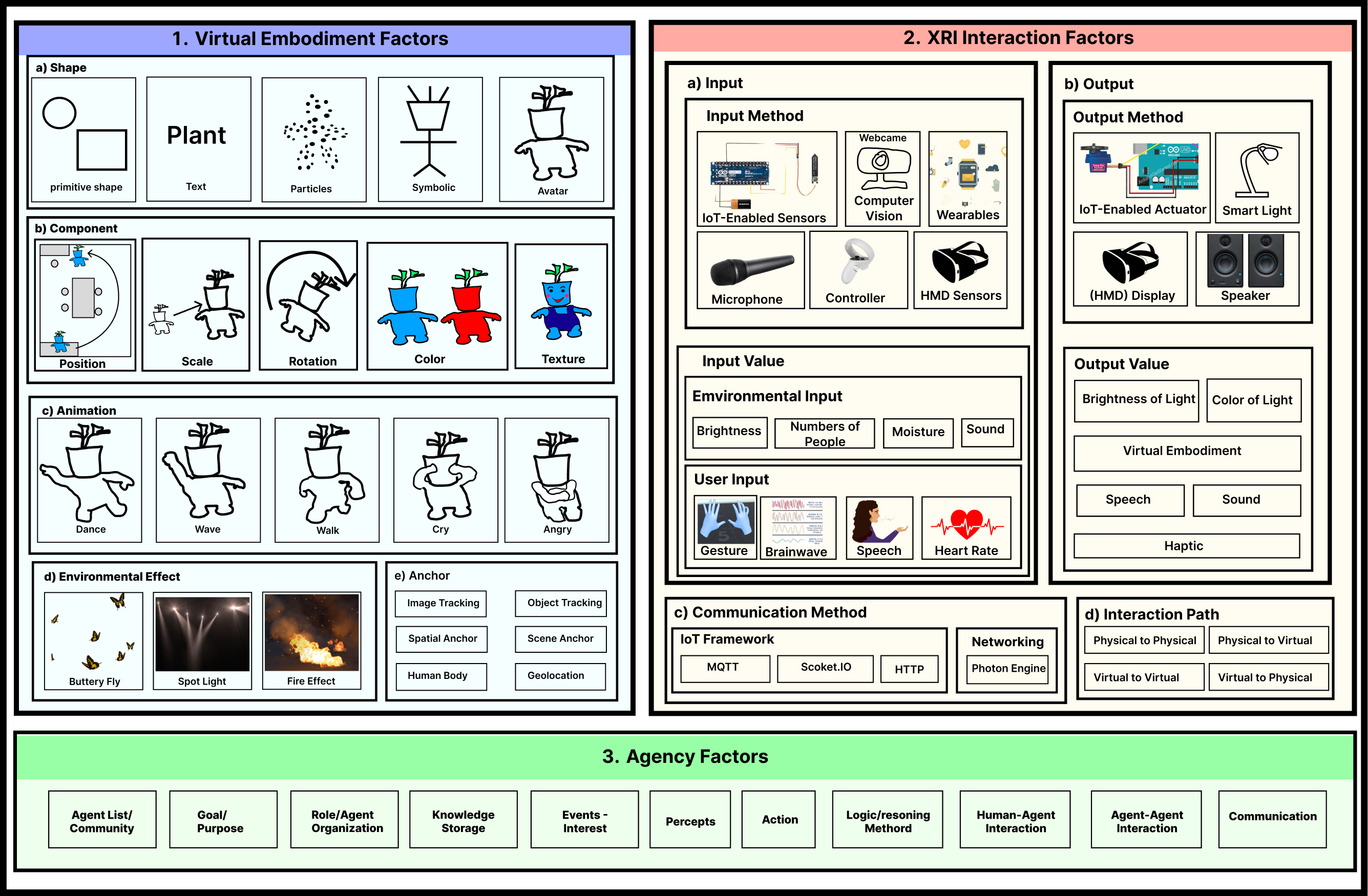}
 \caption{Design elements for XRI Applications with a focus on virtual embodiment methods, XRI interaction sensing components, and agent system designs, as in \cite{holz2011mira} \cite{guan2022extendedbody}.}
 \label{DesignTheory}
\end{figure*}

\begin{figure*}[!htb]
 \centering 
 \includegraphics[width=\linewidth]{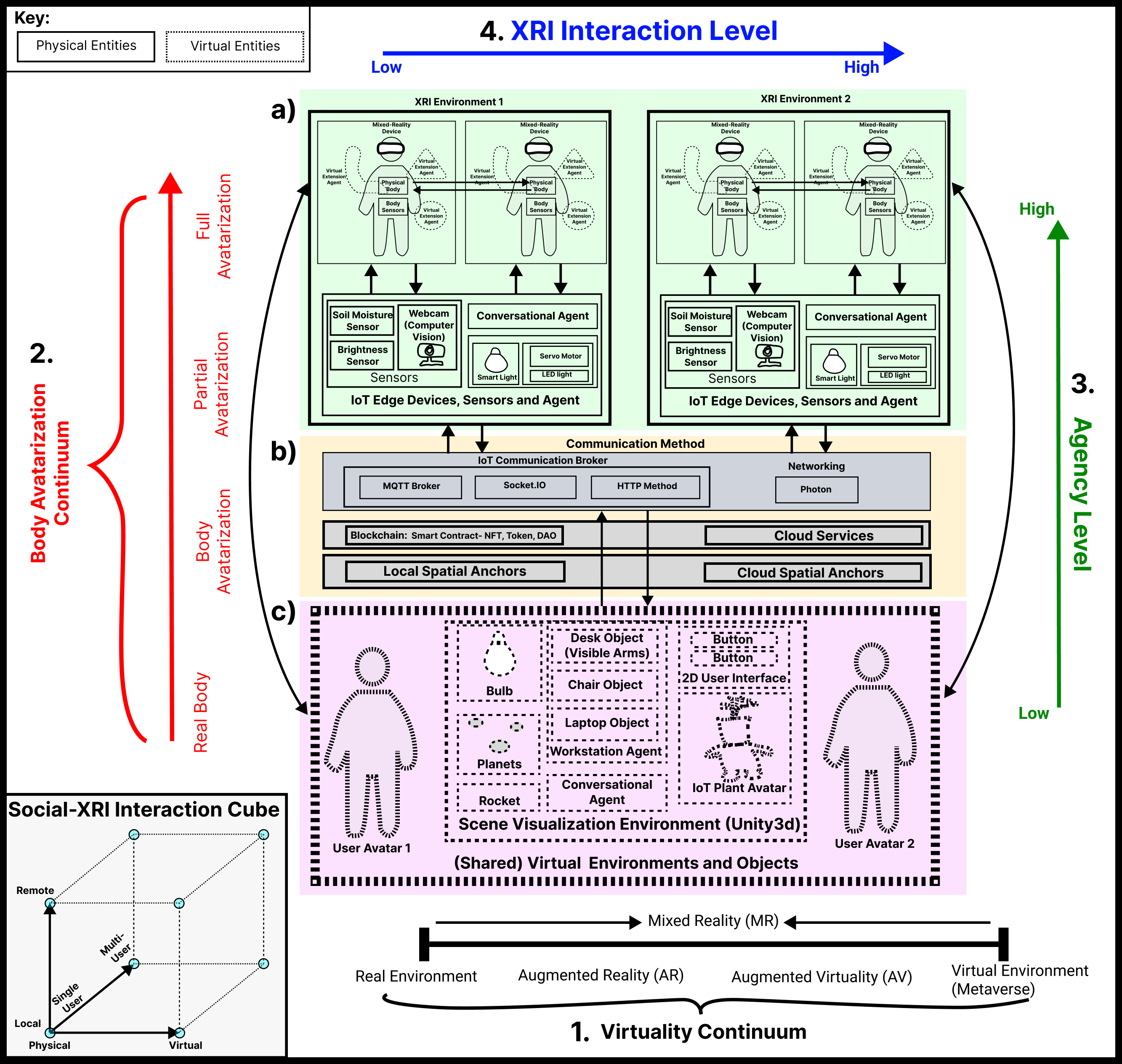}
 \caption{Proposed Social XRI Metaverse architecture for single and multiuser, local and remote\cite{almeida2022telepresence}, physical and virtual interactions. This involves frameworks for XRI interaction\cite{guan2022extendedbody}, level of agency\cite{wooldridge1995agent}\cite{holz2011mira}, body avatarization\cite{guan2022extendedbody}, and level of virtuality across the reality-virtuality continuum\cite{milgram1994taxonomy}. It defines how users in one or more XRI environments can interact, including their IoT edge devices, sensors, and agents, and the communication methods between them that provide access to shared virtual environments and hybrid objects.}
 \label{MultipleUserFramework}
\end{figure*}

Figure \ref{DesignTheory} presents an XRI Agent Design Landscape, with details of the Agency design process, virtual embodiment design method, and XRI interaction with input and output methods, communication method, and interaction path. These design elements are extracted from the authors' previous research on IoT Avatars\cite{morris2020toward}, and Extended Metaverse frameworks\cite{guan2022metabuild}, building on theoretical design frameworks like \cite{holz2011mira}; as a set of three design dimensions: Virtual Embodiment Method, XRI Interaction Method, and Agency. Together, systems that account for these designs would have the foundations for an XRI metaverse experience. These are described as follows:

\subsection{Virtual Embodiment}\label{Embodiment}
For the Virtual Embodiment Method section in Figure \ref{DesignTheory}, it presents an approach toward a more holistic virtual embodiment design process that centres on multiple avatar embodiment and behavior dimensions; this seeks to create digital representations of human or non-human entities that can interact with the physical world. The method described in the paper focuses on the use of shapes to embody parameters that define the appearance and behavior of virtual objects. This includes both static and dynamic aspects of the embodiment, such as  position, scale, rotation, colour, and texture, as well as animations that depict various forms of movement and express different emotions.

\subsection{XRI Interaction} \label{XRIInteraction}
XRI interaction represents virtual-physical input and output information that users and agents in XRI environments send and receive in order to interact with virtual and physical spaces, via the communication method and the interaction path \cite{guan2022extendedbody}. For the input method, it includes IoT-Enabled sensors (Arduino), Webcam to capture context with computer vision models, wearable devices that could be attached to the human body, a microphone for sound and speech detection, and sensors from head-mounted display mixed reality devices and their controllers. With these input methods and devices, the system could capture environmental values such as brightness, the number of people in the space, moisture and sound, and the user input including gesture, brainwave, speech, heart rate, etc.
For the output method, it includes the IoT-Enabled actuator (with Arduino, etc.), smart lights such as Philips Hue, an HMD display, and a speaker. These provide the feedback of brightness and colour of light, display virtual embodiment, speech and sound, and haptic feedback. 
For the communication method, the IoT framework is presented for virtual and physical communication, including MQTT\footnote{https://mqtt.org/ (accessed on 06-February-2023)}, Socket.IO\footnote{https://socket.io/ (accessed on 06-February-2023)}, and HTTP Request\footnote{https://developer.mozilla.org/en-US/docs/Web/HTTP/Methods (accessed on 06-February-2023)}, which are the common protocol for IoT interaction and are being used in the authors' previous projects. In addition, it also has internal logical connections such as virtual object communication within the game engine (Unity), and multiplayer game networking method with Photon\footnote{https://www.photonengine.com/ (accessed on 06-February-2023)}.
The interaction path should also be considered, as in \cite{guan2022extendedbody}, with physical-to-physical, physical-to-virtual, virtual-to-virtual, and virtual-to-physical. 

\subsection{Agency Design Method}
Agent design is related to how the intelligent agents\cite{holz2011mira} interact with the hybrid environments, humans, and each other.  Prometheus methodology is presented in \cite{padgham2002prometheus} for designing and developing intelligent agent systems with inputs (percepts), outputs (actions), and shared data sources. They also indicate that an agent descriptor is needed to show the functionalities, including the name, description, functionalities and who would interact with them. 

\section{A Design Architecture for Social Multi-user XRI Metaverse Environments}\label{DesignArchitecture}

 


Figure \ref{MultipleUserFramework} presents a social XRI metaverse framework, including components of two XRI environments with multi-user, IoT communication broker, and virtual environments and objects that are shared. Alongside the framework, the level of Body Avatarization \cite{genay2022Avatarization}\cite{guan2022extendedbody}, XRI Interaction\cite{guan2022extendedbody},  Agency\cite{holz2011mira}\cite{padgham2002prometheus} and  ``virtuality continuum'' \cite{milgram1994taxonomy} are the factors to be considered when designing the system. This research builds on the multi-user background and related work (in Section \ref{relatedwork}), toward a new theoretical framework that extends and merges multiple physical and virtual environments (including IoT edge devices and IoT-enabled objects, mixed-reality 3D content) for single or multiple users (local or remote, virtual or physical).

\begin{figure}[!htb]
 \centering 
 \includegraphics[width=\linewidth]{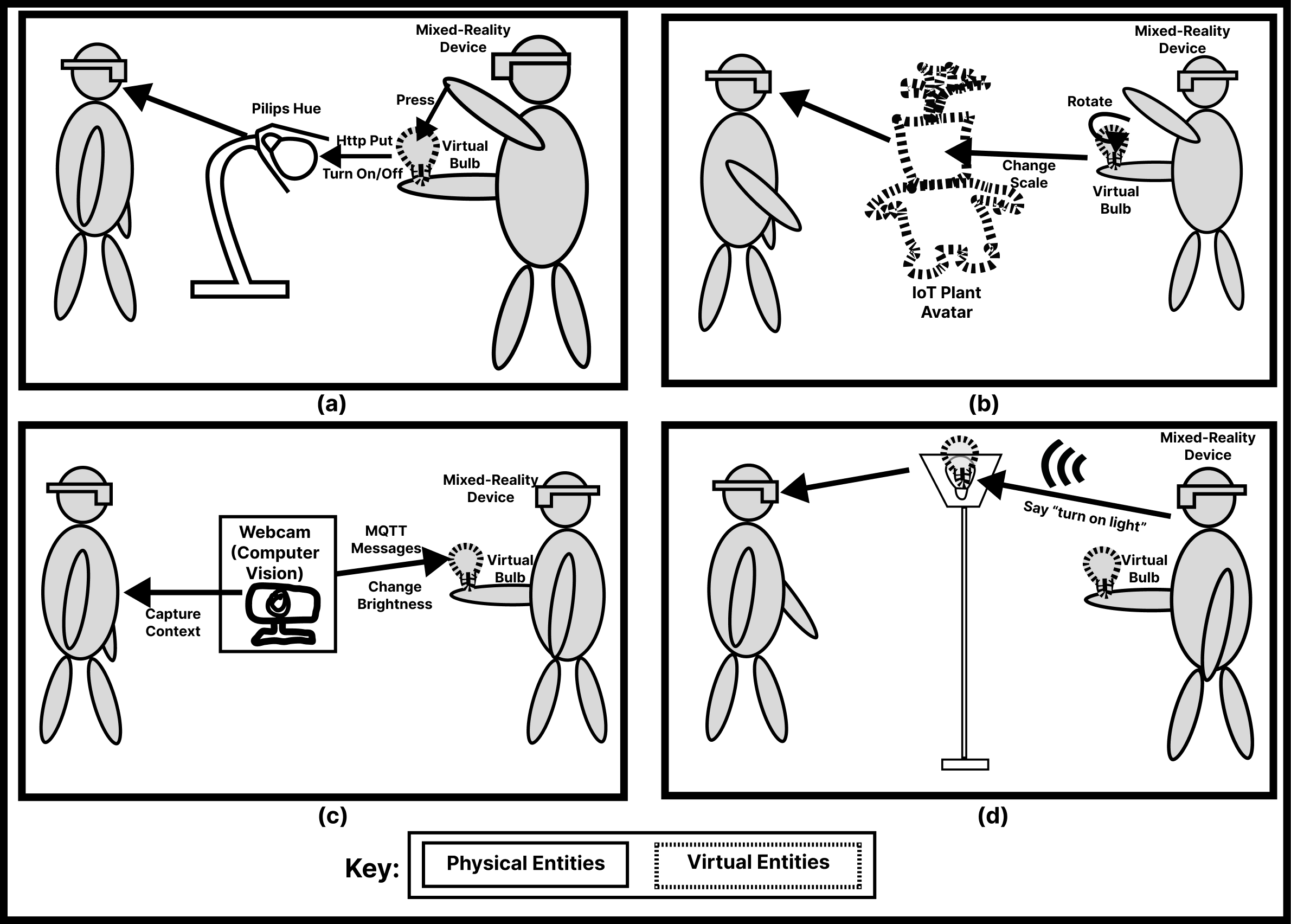}
 \caption{Design scenarios for local multi-user XRI metaverse interaction wherein two users engage with hybrid virtual-physical IoT objects and XRI avatars, as in XRI Environment 1 (see Figure \ref{MultipleUserFramework}). Example interactions include: (a) Manipulating a virtual bulb to turn on/off a physical lamp. (b) Manipulating the scale of a virtual IoT plant avatar. (c) Using computer vision to gather context and interact with virtual agents through physical context changes. (d) Controlling the virtual bulb through conversation. These kinds of interaction are expected to become more common as the metaverse grows in scale and maturity.}
 \label{SameRoom}
\end{figure}


The XRI environments are the local spaces that include IoT-enabled devices to capture context and share through IoT communication broker or be controlled by IoT information. As indicated in the framework, the IoT edge devices could be communicated between the two XRI environments. The virtual environments and objects fit in the ``virtuality continuum'' \cite{milgram1994taxonomy}, as mixed reality objects blended with physical objects in a fully immersive environment(s). IoT enables the ``shared'' values of the virtual content (from the shared hybrid environment objects) that are in communication with the physical environment and which can be accessed and controlled by both the virtual or physical environment elements, as presented in the framework. These hybrid objects could be accessed by the XRI environments virtually, and can be shared for multiple users to interact with, from one individual's local environment, to another individual's remote environment.

\begin{figure*}[!htb]
 \centering 
 \includegraphics[width=\linewidth]{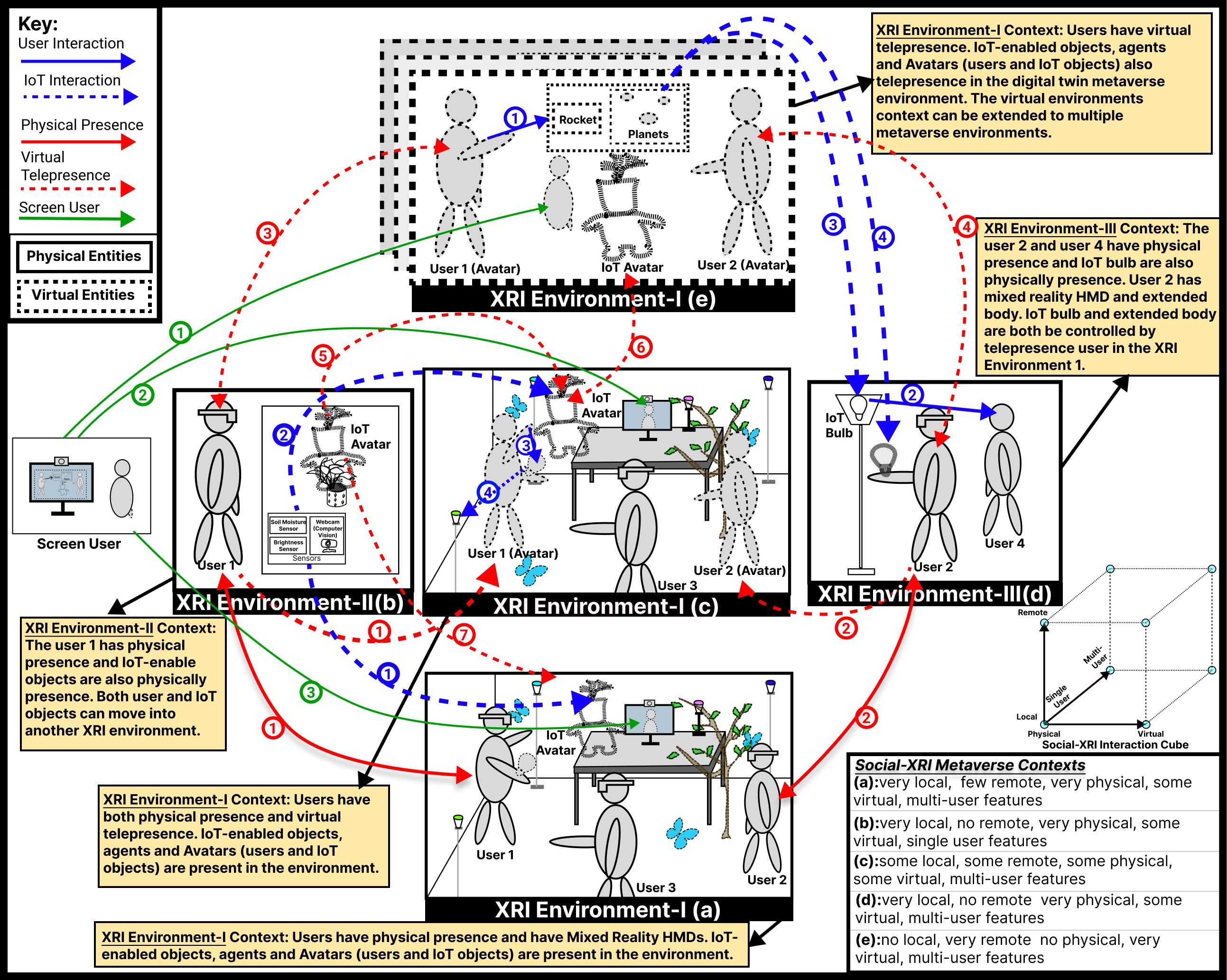}
 \caption{Interactions in the social XRI metaverse are envisioned to scale across multiple environments (see XRI Environment I, II and III), connecting one or more users with diverse sets of IoT-enabled edge devices, agents, and XRI avatars, regardless of their physical or virtual locations, positions across the XRI environment, avatar embodiment, or the access displays used to engage within the metaverse (i.e., screens or HMD's). Example environment configurations are as described above for (a), (b), (c), (d), and (e), as well as the different kinds of interaction (user interaction, IoT interaction, physical presence, virtual telepresence, and traditional display interactions. This level of complex interactions across physical and virtual spaces (single and multiuser, local and remote\cite{almeida2022telepresence}, physical and virtual interactions) must be addressed by social XRI metaverse systems. The Social-XRI Interaction Cube (bottom right) shows multiple dimensions and highlights where the context situations (a),(b),(c),(d), and (e) fit within the dimensions.}
 \label{HyperScenario}
\end{figure*}


In terms of the users, in one XRI environment, they could interact with each other through the physical-to-physical path (the normal way to interact in the real world), the physical-to-virtual path with virtual objects or partial avatarization body\cite{genay2022Avatarization}\cite{guan2022extendedbody}, the virtual-to-physical path with their partial avatarization to manipulate the physical objects and environments, and virtual-to-virtual path with partial avatarization body or full avatarization body in the full virtual environment or remote XRI environment to interact with virtual objects.

In terms of the level of agency\cite{holz2011mira}, this refers to the ability of intelligent actors to interact with users and each other, and also control the hybrid objects in the XRI environment. In this sense, conversational agents are common, as IoT interfaces, and in this case a conversational agent is considered as part of the agency considerations to provide chat with users, which means it could be aware of the speech of the users, understand the speech, and provide feedback (with the speech in the speaker or virtual embodiment). One of the recent popular examples is ChatGPT\footnote{https://openai.com/blog/chatgpt/ (accessed on 07-February-2023)} that could interact with users through text, and it could be potentially used in the Social XRI metaverse environment with the virtual embodiment.

\section{Social Interaction Scenarios in XRI Multi-user Metaverse}\label{SocialInteractionScenarios}

\subsection{Social XRI Multi-user Metaverse Hybrid Interactions in Local Environments}

Figure \ref{SameRoom} focuses on the design of scenarios for multi-user interactions in the same room within a hyper-connected metaverse environment. The interactions involve virtual-to-physical, virtual-to-virtual, physical-to-virtual, and physical-to-physical paths and involve the use of XR-IoT hybrid bodies and virtual extensions. Figure \ref{SameRoom}(a) In the first scenario, an XR-IoT hybrid body interacts with a virtual bulb as a body extension by pressing it to turn on or off a physical lamp through HTTP PUT, which can be observed by others and is considered a virtual-to-physical path interaction. Figure \ref{SameRoom}(b) Another scenario involves an XRI-hybrid body manipulating the scale of a virtual IoT plant avatar using its virtual bulb extension, which can be observed by others virtually and is considered a virtual-to-virtual path interaction. Figure \ref{SameRoom}(c) A physical context is affected by other users, captured by a webcam with computer vision, to affect the virtual extension agent (virtual bulb) on the XR-IoT hybrid body, which is considered a physical-to-virtual path interaction. Figure \ref{SameRoom}(d) A scenario addresses control of the virtual bulb through conversation from a user, which can be observed by others and is considered a physical-to-physical path interaction.

\subsection{Social XRI Multi-user Metaverse Hybrid Interactions across Local and Remote Environments}


Figure \ref{HyperScenario} presents the highly connected multi-user metaverse scenarios with the seamless integration of virtual and physical spaces. These spaces can be described by the Social-XRI Interaction Cube (bottom right), based on physical-virtual, local-remote, single user and multi-user dimensions, for each of the context situations in XRI-Environments (a),(b),(c),(d), and (e).

The solid blue arrow represents the direct interaction of the user with virtual or physical objects. For example, as indicated in number 1 of the solid blue arrow,  the user has a full avatar in an immersive metaverse environment to interact with the virtual rocket and planets (virtual-to-virtual interaction). At the same time, the dashed blue arrow symbolizes the communication of IoT information between objects, with bi-directional virtual-to-physical interaction, such as the number 1 of the dashed blue arrow indicates the virtual IoT avatar is controlled by the physical sensors in XRI Environment-II(b) Context. 

The solid red arrow signifies movement between physical spaces; as indicated in solid red arrow number 1, user one could move physically between the XRI Environment-II(b) Context and XRI Environment-I(a) Context. In addition, the dashed red arrow represents virtual telepresence in a mixed-reality environment or an immersive metaverse (VR) environment. On the one hand, for user telepresence, as indicated in the dashed red arrow number 2, User 2 in XRI Environment-III(d) Context could have a virtual presence in the XRI Environment-I(c) Context that could be viewed by User 3 (physical presence) or in the XRI Environment-I(e) Context that could be viewed by another virtual avatar (such as User 3).

The solid green arrow signifies the use of traditional devices such as a laptop with a screen by the user, providing a more conventional means of interaction within the virtual environment. The numbers 2 and 3 of the green arrows are the most traditional in a meeting through webcam and screen presence, like Zoom\footnote{https://zoom.us/ (accessed on 07-February-2023)} and Teams\footnote{https://www.microsoft.com/microsoft-teams/ (accessed on 07-February-2023)}. Number 1 of the green arrow using screen devices to access the three-dimensional metaverse environment (XRI Environment-I(e) Context) and present as an avatar that the mouse and keyboard could control, such as Decentraland and Spatial (see Table \ref{Comparison}).

\subsection{Social XRI Multi-user Metaverse Use-Case Scenarios across Local and Remote Environments}
In a social XRI metaverse research lab scenario, the lab's room could be considered a physical and hybrid XRI environment where researchers and students work in person or remotely. The User 1, 2, and 3 are physically working in the lab as indicated in the XRI Environment-I(a) Context, and they could experience the XRI environment with a virtual IoT Avatar, trees, butterflies, and physical light that could be controlled with the virtual objects\cite{morris2020toward}\cite{morris2021xri}\cite{guan2022metabuild}\cite{guan2022thesis}. 

The IoT Avatar in XRI Environment-I(a) Context is the virtual telepresence of the plant (indicated by red dashed arrow number 7) and its IoT information embodiment (indicated by blue dashed arrow number 1). User 1 could be aware of the status (emotion) \cite{morris2020toward} of the physical plant in XRI Environment-II(b) Context. If the plant is sad and needs to be watered, User 1 could physically move back to XRI Environment-II(b) Context to water the plant to make it happy. During the commute from Physical XRI Environment to XRI Environment-II(b) Context or in XRI Environment-II(b) Context, User 1 is still able to have virtual telepresence as a full avatar to the XRI Environment-I(a) Context (as indicated in red dashed arrow number 1) to communicate with User 2 and 3. At the same time, User 1 in XRI Environment-I(c) Context as an Extended-XRI body with a virtual light bulb extension, the virtual light bulb attached to the virtual avatar hand could be pressed (as indicated in solid blue arrow number 3) and control the physical light (as indicated in blue dashed arrow number 4). Such a scenario could apply when the users attempt to have an immersive meeting and provide a better lighting environment remotely.

In terms of the XRI Environment-I(e) Context, it is a completely virtual environment that users need to access through a VR headset or screen-based devices. Users 1 and 2 are virtually present in that space (as indicated in red dashed arrow numbers 3 and 4) and embodied as full avatars. They could host meetings, play games and etc. in the place. If User 2 would like to switch back to XRI Environment-II(b) Context, User 1 as an avatar in the XRI Environment-I(e) Context could still connect with User 2, through interaction with the virtual objects in the shared virtual environment, i.e., rocket and planet, (as indicated in solid blue arrow number 1) and control the physical lights (as indicated in blue dashed arrow number 3) which could also be saw by User 4 in the XRI Environment-III(d) Context (as indicated in solid blue arrow number 2) or the virtual bulb extended body attached to User 2 (as indicated in blue dashed arrow number 4). 

\subsection{Building the Social-XRI Metaverse: Technologies and Implementation Considerations}


The above Social XRI Metaverse scenarios (Figure \ref{HyperScenario}) can be realized with existing technologies (as in Figure \ref{MultipleUserFramework}). These include approaches to provide avatar engagement and expression and telepresence in digital twin metaverse environments, based on: environment and body sensors; webcams; conversation agent controllers; smart lights and edge devices; actuators; communication channels; networks; information brokers (e.g., via blockchains); cloud services; mixed reality content anchor systems; HMDs; wearable devices; trackers; feedback devices (e.g., haptics); together with traditional displays, and AI Frameworks. 

These technologies provide the core elements needed for social and multiuser and multi-agent telepresence, however, in terms of avatar design, the framework can incorporate existing avatar tools, such as 
the avatar creation system from Meta\footnote{https://www.theverge.com/2021/4/23/22398060/oculus-new-avatars-editor-features-vr-virtual-reality-facebook-quest-rift (accessed on 15-April-2023)}, 
and Ready Player Me\footnote{https://readyplayer.me/(accessed on 15-April-2023)},
The emotion and expressiveness of users within these environments can be incorporated by avatar by face tracking sensors such as those of current generation HMD devices (e.g., Meta Quest Pro). Avatar movement is commonly based on inverse kinematics (IK), based on the position of the two controllers and the HMD position\footnote{https://www.uploadvr.com/meta-quest-2-body-tracking-without-trackers/ (accessed on 15-April-2023)}, or via on-body trackers like the HTC Vive Tracker\footnote{https://www.vive.com/ca/accessory/tracker3/ (accessed on 15-April-2023)}, and/or external vision-based  trackers (e.g., webcam computer vision models such as PoseNet\footnote{https://blog.tensorflow.org/2018/05/real-time-human-pose-estimation-in.html (accessed on 15-April-2023)}) or depth camera tracking (such as Kinect\footnote{https://learn.microsoft.com/en-us/azure/kinect-dk/body-joints (accessed on 15-April-2023)}). Further, volumetric capture is a viable possibility to capture the user moving in real-time with a depth camera and to produce the volume of video in the remote space, such as Holoportation\footnote{https://www.microsoft.com/en-us/research/project/holoportation-3/ (accessed on 15-April-2023)}.

\section{Discussion}\label{Discussion}

The proposed framework presents a perspective on the future of the metaverse with social and XRI components for multiuser and multiagent interactions and experiences. This approach can bring multiple benefits to the metaverse platform, however, it also has challenges to address, as seen in \cite{lee2021all} across both the underlying platform technologies and the resulting ecosystem.
In terms of benefits social XRI metaverse ecosystems may help with: \emph{Remote-work and Co-working} -- The metaverse's ability to integrate multiple user environments and modalities facilitates a digital workspace where real-time collaboration can mimic in-person interactions regardless of physical geographical constraints. \emph{Metaverse Connectedness} -- The proposed design frameworks may help with reducing the inherent task-switching disruptions that occur when users engage within and across multiple virtual and physical environments simultaneously \cite{guan2022metabuild}, as it streamlines both physical and virtual spaces into a more cohesive and holistic meta-space for user interaction.
\emph{Multi-agent Interaction} -- The designs indicate the possible incorporation of both human users as well as non-human agent components, which can both express and engage across the virtual-physical, local-remote dimensions. This encourages the design toward a new form of hyper-connected multi-agent telepresence, and potential for new forms of human-environment interaction, human-agent, and human-human interactions.

On the other hand, in terms of challenges, these include: \emph{Integration Complexity} -- The creation of a seamless and immersive metaverse is a complex endeavor, requiring significant advances in areas like artificial intelligence, mixed reality, and IoT technologies. These technologies each require specialized frameworks that developers need to merge into a single runtime framework. The frameworks shown in this work highlight some of the many sides of this issue. \emph{Privacy and data security}-- As a digital twin of the real world, the metaverse will likely process substantial personal data, raising significant concerns about privacy and data security, as well as ownership of information. This is an open area of research and must account for this from multiple sides of the social XRI problem. \emph{Latency} -- Ensuring real-time interactions in a hyper-connected meta-environment is a considerable challenge, especially when the systems involved are decentralized, potentially large scale, and also involve high amounts of graphical information, image data, and other artificial intelligence data and models. Such hyper-connectedness is required, but brings with it a heavy data management cost. As a result, high latency could disrupt the user experience, particularly in time-sensitive activities.
Although the framework proposed does not address these challenges currently, it is expected that future advances in computation and latency will enable these new forms of multi-user interaction within the social XRI metaverse to be achievable.

\section{Summary}\label{Summary}

This work has presented an exploration of how the metaverse concept of a digital twin overlaying the physical environment can be made more extended through the use of XRI technologies, toward a social XRI metaverse. 
The design considerations for this have been presented, first for XRI systems broadly, and have been extended into a design architecture for multi-user social interactions. This offers a path toward frameworks for creating such metaverse environments. Further, key scenarios related to  the types of user-user or user-agent or even agent-agent interactions across these systems have been identified as an area for exploring social-metaverse concepts. 

The outcomes of this work set the stage for new prototype concepts, and a testbed for examining the benefits and limitations of social interactions within such a hyper-connected hybrid virtual-physical shared environment. In particular, future research will evaluate and test the architecture and the experiences shown in this work, across the dimensions identified. 
Likewise, it is worth highlighting that the design challenge within the social XRI metaverse is complex, and a single architectural framework may not be able to be instantiated. 
However, approaches involving prototyping and proof-of-concept development
can be used to explore this concept and to consider the range of human factors involved \cite{vicente2013human}.
It remains for future metaverse researchers, developers, designers, and creators to engage in bringing this concept forward, to examine how to better accommodate shared social and multiuser real-world XRI metaverse experiences and interactions across the entirety of the reality-virtuality spectrum.

\section*{Acknowledgment}
This work was supported by funding from the Tri-council of Canada under the Canada Research Chairs program.

\bibliographystyle{IEEEtran}
\bibliography{references}

\begin{IEEEbiographynophoto}{Jie Guan}
is an MFA graduate of the Digital Futures program at OCAD University, with a background in Digital Painting and Expanded Animation from the same institution. With a focus on extending the Metaverse, he combines Internet of Things and Extended Reality to push the boundaries of immersive experiences.
\end{IEEEbiographynophoto}

\begin{IEEEbiographynophoto}{Alexis Morris}
is the Canada Research Chair (Tier II) in the Internet of Things, Director of the Adaptive Context Environments Lab, and Associate Professor at OCAD University. He is a Computer Scientist with a Ph.D., from the University of New Brunswick, specializing in Metaverse mixed reality adaptive interfaces and context-aware smart environments.

\end{IEEEbiographynophoto}

\end{document}